\documentstyle[preprint,aps,floats,epsfig]{revtex}
\tighten

\newcommand{\Lav}{\mbox{$\langle L \rangle$}}
\newcommand{\grow}{\mbox{$\alpha$}}
\newcommand{\gammaeff}{\mbox{$\gamma_{\hbox {\rm eff}}$}}
\newcommand{\Lstar}{\mbox{$L^{*}$}}
\newcommand{\phistar}{\mbox{$\phi^{*}$}}
\newcommand{\xiL}{\mbox{$\Xi$}}

\newcommand{\Ree}{\mbox{$\left< R_e^2 \right>$}}
\newcommand{\Rgg}{\mbox{$\left< R_g^2 \right>$}}
\newcommand{\Rs}{\mbox{$R_0$}}

\begin{document}

\title{Computational confirmation of scaling predictions\\ for equilibrium polymers}

\author{J.~P.~Wittmer$^1$\thanks{E-mail: j.wittmer@ed.ac.uk},
A.~Milchev$^2$, M.~E.~Cates$^1$}
\vskip 2.0truecm
\address{
$^1$Department of Physics and Astronomy, University of Edinburgh,\\
JCMB King's Buildings, Mayfield Road, Edinburgh EH9 3JZ, UK\\
\vskip 0.3truecm
$^2$Institute for Physical Chemistry, Bulgarian Academy of Science,
1113 Sofia, Bulgaria}

\setcounter{page}{0}
\date{PACS numbers: 61.25.Hq, 05.40+j, 05.50.+q, 64.60.Cn, 82.35.+t, 36.20-r}
\maketitle

\begin{abstract}
We report the results of extensive Dynamic Monte Carlo simulations of
systems of self-assembled Equilibrium Polymers without rings in good solvent.
Confirming recent theoretical predictions,
the mean-chain length is found to scale as
$\Lav = \Lstar \left(\phi/\phistar \right)^{\grow}
\propto \phi^{\grow} \exp(\delta E)$
with exponents $\grow_d=\delta_d=1/(1+\gamma) \approx 0.46$
and $\grow_s = [1+(\gamma-1)/(\nu d -1)]/2 \approx 0.60, \delta_s=1/2$
in the dilute and semi-dilute limits respectively.
The average size of the micelles, as measured by the
end-to-end distance and the radius of gyration, follows a very similar
crossover scaling to that of conventional quenched polymer chains.
In the semi-dilute regime, the chain size distribution is found to be
exponential, crossing over to a Schultz-Zimm type
distribution in the dilute limit. The very large size of our simulations
(which involve mean chain lengths up to $5000$, even at high polymer densities)
allows also an accurate
determination of the self-avoiding walk susceptibility exponent 
$\gamma = 1.165 \pm 0.01$.
\end{abstract}

\thispagestyle{empty}

\newpage

Systems in which polymerization is believed to take place under
condition of chemical equilibrium between the polymers and their
respective monomers are termed ``equilibrium polymers'' (EP).
Examples include liquid sulfur\cite{Scott,WheelerPfeuty80} and
selenium\cite{Faivre}, poly-$\alpha$-methylstyrene\cite{Greer},
giant micelles\cite{CatesCandau} and protein filaments\cite{Proteins}.
In several of these systems, the behaviour is strongly affected by the
presence of polymeric rings\cite{rings}. For reasons that are not
entirely clear however, ring-formation effects seem to be negligible in other
cases, including that of giant micelles\cite{CatesCandau}. Since
the latter are among the most widely studied examples of equilibrium
polymers, the ring-free case is of some interest. For micelles there
has been controversy concerning the dependence of the mean chain length
on volume fraction, described by the growth exponent $\alpha$.
Indirect evidence, based on viscosity measurements interpreted via
the reptation-reaction model\cite{CatesCandau}, suggests in some semidilute
systems
a value of $\alpha\simeq 1.2$ different\cite{Schurtenberger} from the
prediction of scaling theory $\alpha\simeq 0.6$ \cite{Cates88,Schaefer,Schoot}.

In this communication we present results for EP without rings,
derived by means of a new efficient Dynamic Monte Carlo algorithm\cite{WMC},
sampling over a wide range of volume fractions $\phi$ and
scission energies $E$. (This is the energy to create a pair of chain ends.)
We confirm unambiguously the scaling expectations based on the classical
behaviour of conventional quenched polymers \cite{Cates88,Schaefer,Schoot}.
The crossover scaling of both mean chain length \Lav\ and mean size of the
chains are discussed.
Our results provide strong support for the form of the Molecular Weight
Distribution (MWD), $c(L)$, predicted on the ground of renormalizational
analysis\cite{Schaefer,Schoot}, and discard an earlier conjecture\cite{Guj}.
In passing, we obtain an accurate estimation for the susceptibility
exponent $\gamma$ of self-avoiding walks, which can be directly
measured from the slope of $c(L)$ in semi-log coordinates.

At the level of a Flory-Huggins Mean Field Approximation (MFA) the
grand potential $\Omega$ for a system of EP may be written, in units of
$k_BT$, as
\begin{equation}
\Omega =\sum_{L} c(L)b^d \left[ \ln\left(c(L)b^d \right) + E + L \mu\right].
\label{eq:FMFA}
\end{equation}
The first term is the entropy of mixing, the second the scission energy $E$
of a bond and the third entails the usual Lagrange multiplier $\mu$ for the
constraint of constant total density of the monomers
$\phi = \sum_{L}^{\infty} L c(L)$ (in suitable units).
For convenience the persistence length $l_p$ is set equal to the size of a
monomer $b$.
Without loss of generality we have set to zero in Eq.~(\ref{eq:F})
the part of the free energy linear in chain length.
It is useful to introduce the normalized probability distribution
$p(L) \equiv \left(\Lav/\phi\right) c(L)$, so that $\sum_{L} p(L) = 1$.
Minimizing with respect to the MWD, subject to the constraint,
yields an exponentially decaying MWD, with $p(L) dL = p(s)ds = \exp(-s) ds$,
where the chemical potential sets the scaling variable $s=\mu L$.
This distribution requires $\Lav \mu =1$, i.e. the scaling variable $s$
is given, in the MFA, by the reduced chain length $x\equiv L/\Lav = s$.
The mean chain length is found to be $\Lav \propto \phi^{1/2} \exp(E/2)$.
This result is expected to be a good approximation around the
$\theta$-temperature of the system (not of interest here) and also in the
dense regime where excluded volume interactions are largely screened, 
and the chains behave like (gaussian) random walks\cite{YA1}.
The MFA coincides, in the low concentration limit, with the law of mass action.

One can easily extend the above analysis to dilute and semi-dilute
solutions of EP. We recall\cite{deGennes79} from ordinary polymers that
the correlation length \xiL\ for a chain of length $L$ in the dilute limit
is given by the size, $R$, of the chain: $\xiL = R \propto L^{\nu}$.
When chains become so long that they start to overlap at
$L \approx \Lstar \propto \phi^{-1/(\nu d -1)}$,
the correlation length of the chain levels off and reflects the
(chain-length independent) `blob-size', $\xi$: thus $\xiL = \xi \propto
\Lstar^{\nu}$.  Here $d$ is the dimension of space and
$\nu \approx 0.588$ is the swollen chain exponent.
The introduction of these correlations into the picture is relatively
straightforward. The mean-field approach remains valid~\cite{Cates88}
as long as the basic `monomer' is replaced by
coarse-grained blobs of monomers, of size \xiL. Accordingly we may write
the grand potential as
\begin{equation}
\Omega = \sum_{L} c(L) \xiL^d \left[ \ln\left( c(L) \xiL^d \right)
    - \ln \left(\xiL/b\right)^{d+\theta} + E + L\mu \right].
\label{eq:F}
\end{equation}
In Eq.(\ref{eq:F}) the second term in the brackets reflects the
{\em gain in entropy} when a chain breaks so that the two new ends can
explore a volume $\xiL^d$. Entropy is increased because the excluded volume
repulsion on scales less than \xiL\ is reduced by breaking the chain;
this effect is accounted for by the additional exponent $\theta=(\gamma-1)/\nu$.
Note that in $3d$ one has $\theta \approx 0.3$ with $\gamma \approx 1.17$.
In MFA $\gamma=1,\;\theta=0$, and Eq.~(\ref{eq:F}) simplifies to
Eq.~(\ref{eq:FMFA}), where correlations, brought about by the mutual
avoidance of the chains, i.e. excluded volume, are ignored.
In the dilute regime Eq.~(\ref{eq:F}) can be rewritten as
$\Omega =\sum_{L} c(L) \left[ \ln c(L) + (\gamma-1) \ln L + E + L \mu\right]$,
so that the relation to the well known partition function of self-avoiding
walks\cite{deGennes79} (with an effective coordination number $\tilde{z}$),
$Q_L \propto \tilde{z}^L L^{\gamma-1}$, is evident. Hence Eq.(\ref{eq:F})
covers the entire concentration range.

Minimizing Eq.~(\ref{eq:F}) at fixed $\phi$, yields the MWD
$p(L) dL = p(s) ds \propto \xiL(s) \exp(-s) ds$.
Note that for large chain length $L$ we have
$p(L)dL \propto \exp(-\gammaeff x) dx$.
Here, we have introduced the effective exponent
$\gammaeff \equiv s/x = \Lav \mu= \left< s \right>$. Since in general
$\gammaeff$ is not unity, the scaling variable $s$ does not equal the
reduced chain length $x$.
In the two limits, far away from the crossover line dividing dilute from
semidilute behavior,
one can readily
calculate  the integrals. In the semi-dilute limit we obtain $\gammaeff =
1$, so that in this case $s=x$ as in conventional MFA. In the dilute
limit, however, $\gammaeff = \gamma$, so that $s=\gamma x$.
Substituting these results, the distributions become
\begin{equation}
p(x) dx =
\left\{ \begin{array}{ll}
\exp(-x) dx & \mbox{($\Lav \gg \Lstar$)} \\
\frac{\gamma^{\gamma}}{\Gamma(\gamma)} x^{\gamma-1} \exp(-\gamma x) dx
        &\mbox{($\Lav \ll \Lstar$)}
\end{array}
\right.
\label{eq:p}
\end{equation}
Using again the conservation law one obtains a mean chain length
\begin{equation}
\Lav = \Lstar (\phi/\phistar)^{\grow} \propto \phi^{\grow} \exp(\delta E)
\label{eq:Lscal}
\end{equation}
with exponents
$\grow_d=\delta_d=1/(1+\gamma)\approx 0.46$ in the dilute and
$\grow_s=1/2 (1+(\gamma-1)/(\nu d-1) \approx 0.6$, $\delta_s =1/2$ in the
semi-dilute regime.
Note that for self-consistency  we require $\Lav_d=\Lav_s$ at $\phi=\phistar$,
which imposes that
the crossover density obey $\phistar \propto  \exp(-E/\varphi)$ and the
crossover chain length $\Lstar \propto  \exp(E/\kappa)$.
Here the exponents are $\varphi= (\grow_s-\grow_d)/(\delta_s-\delta_d)
\approx 3.8$
and $\kappa= (\nu d -1 ) \varphi \approx 2.93$;
similar relations hold between the amplitudes.

\begin{figure}
\centerline{
\epsfig{file=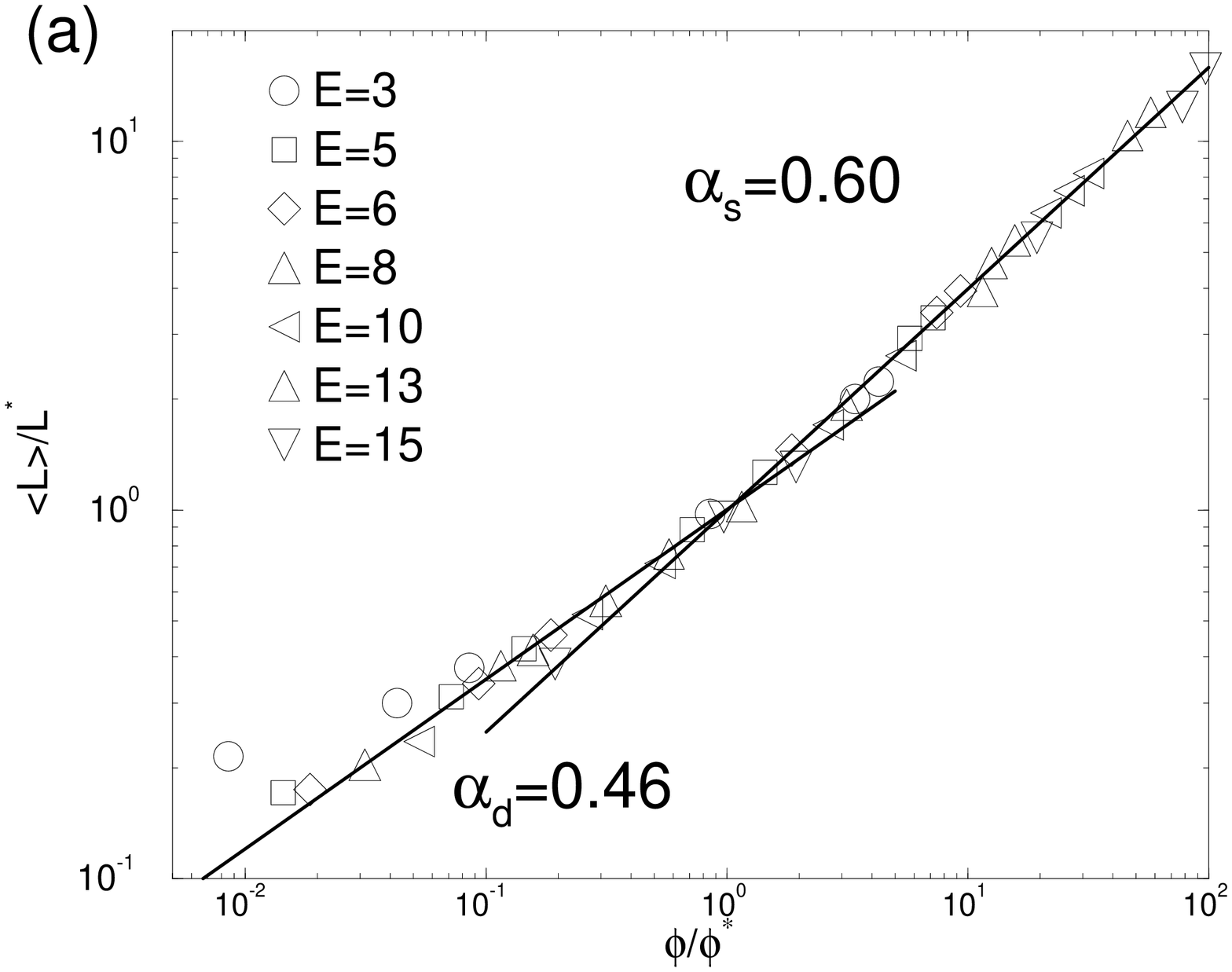,width=70mm,height=80mm,angle=-90}
\epsfig{file=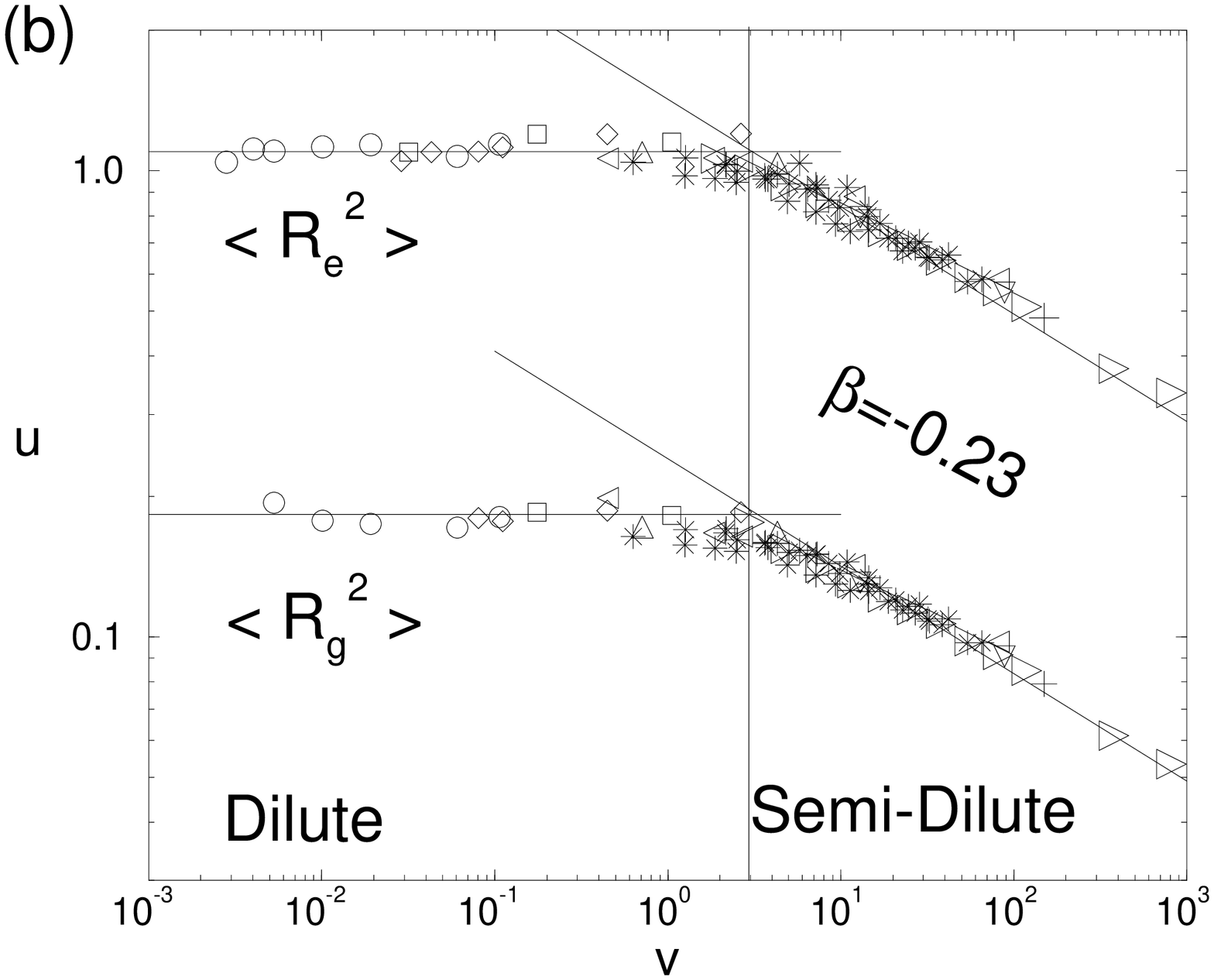,width=70mm,height=80mm,angle=-90}
}
\caption{Crossover scaling versus reduced density $\phi/\phistar$ for
(a) rescaled mean chain length $\Lav/\Lstar$ confirming
scaling Eq.~(\ref{eq:Lscal}) and
(b) reduced mean-square chain size $u$
(Same symbols as in (a), stars for dead chains).
\label{fig:Cross}}
\end{figure}

We have confirmed all the major predictions of this scaling analysis by means
of an efficient new Dynamic Monte Carlo simulation, based on a modified
bond fluctuation algorithm for polymer chains\cite{Wolferl}:
in EP systems bonds between monomers break and recombine permanently
and the {\em chains} are only transient objects;
the (two saturated or unsaturated) {\em bonds} of each monomer become the 
principle objects of the data structure~\cite{WMC}. 
We were able to obtain mean chain lengths of up to $\Lav \approx 5000$
with large system sizes (e.g. we have $62,500$ monomers at $\phi=0.5$).
These values are large enough for meaningful comparison with experimental
systems of giant micelles,
and represent at least an order of magnitude increase over those studied
previously~\cite{Kroeger}.
(They also represent higher mean chain lengths at melt densities than any 
we have so far seen published for conventional quenched chains.) 
Details of the method, and more of our new results, will be presented 
elsewhere \cite{WMC}. Note that our simulation is limited by
system size rather than equilibration time: 
we are unable to go to higher $E$ (lower temperatures) to increase
$\Lav$ further because the largest chain present would then contain too
high a proportion of all the monomers present. 
This leads to finite size corrections, studied in detail in Ref.~\cite{WMC}.

The growth exponents $\grow$ for dilute and semidilute regimes are most readily
analysed by plotting
the number of blobs $\Lav/\Lstar$ versus the reduced density $\phi/\phistar$.
(This is suggested by the scaling prediction Eq.~(\ref{eq:Lscal}).)
The data are found to collapse onto a single master curve in
Fig.~\ref{fig:Cross}a with
remarkable precision; figure represents one of the main results of
this work.
The two limiting slopes compare very well with the two predicted growth
exponents
($\grow_d = 0.46$, $\grow_s \approx 0.60$).
In addition, we are able to confirm the other exponents mentioned above
and to obtain all the corresponding amplitude factors~\cite{WMC}.

We show in Fig.~\ref{fig:Cross}b the
crossover scaling plot of the mean chain size as measured from the
end-to-end distance \Ree\ and the radius of gyration \Rgg.
Defining a reference chain size $\Rs = b \Lav^{\nu}$, we plot in
Fig.~\ref{fig:Cross}b the reduced average chain size,
$u_e=\Ree/\Rs^2$ and $u_g=\Rgg/\Rs^2$,
defined via the end-to-end distance and the radius of gyration respectively, 
as functions of a scaling variable $v \propto \phi/\phistar$.
The curves for EP are compared to those of conventional quenched
monodisperse polymer
(so that $c(L) = \delta(L-N)$). The latter data, represented as stars in
Fig.~\ref{fig:Cross}b, was taken from Paul et al~\cite{Wolferl}.
In the semi-dilute limit the chains are Gaussian on length scales larger
than the blob size $\xi$ implying the scaling $u \propto v^{-\beta}$
with exponent $\beta=(2\nu-1)/(3\nu - 1) \approx -0.23$.
It is remarkable that, in their scaling behaviour, EP and dead polymers are
nearly indistinguishable; the two universal scaling functions coincide to
numerical accuracy.

\begin{figure}
\centerline{
\epsfig{file=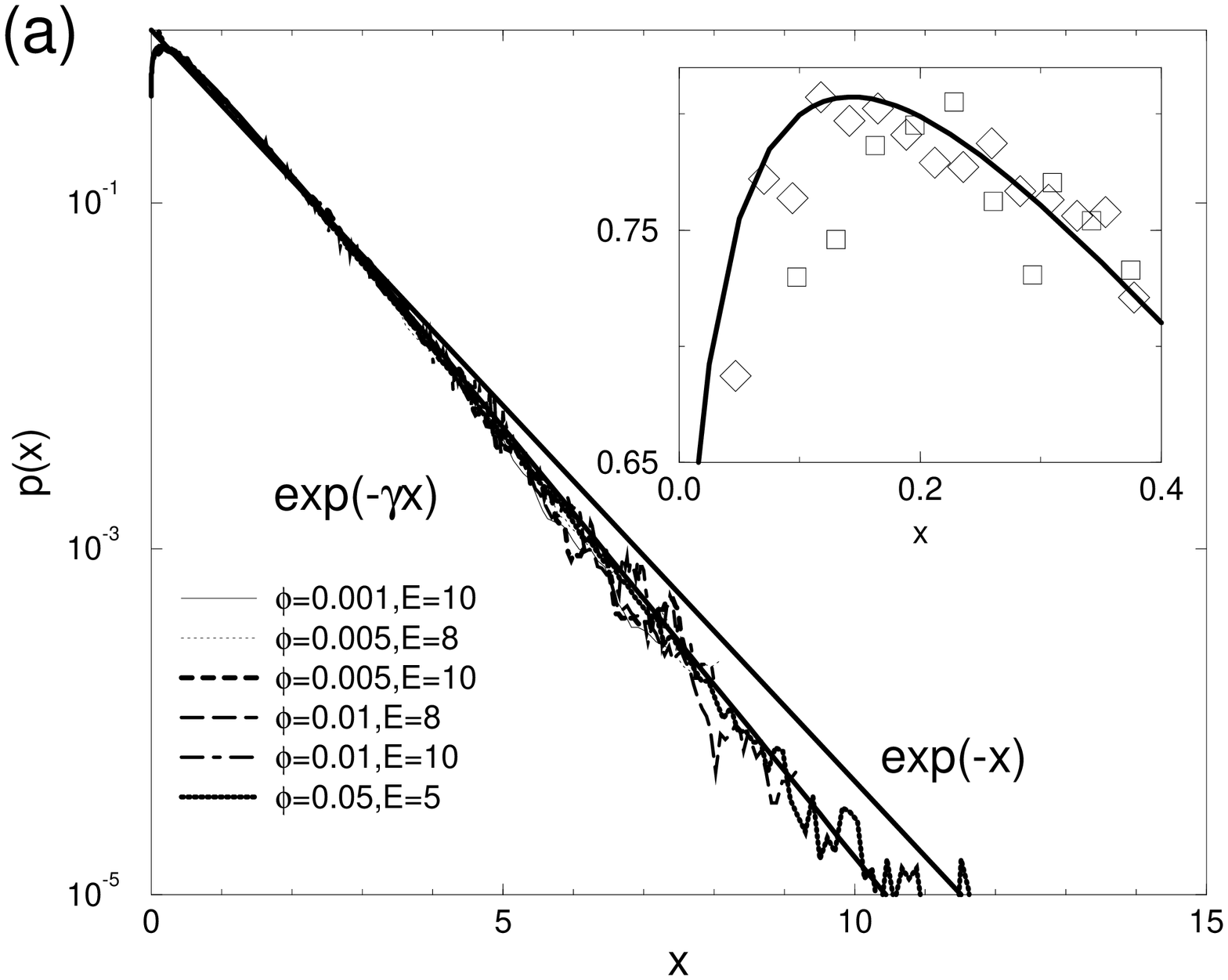,width=70mm,height=80mm,angle=-90}
\epsfig{file=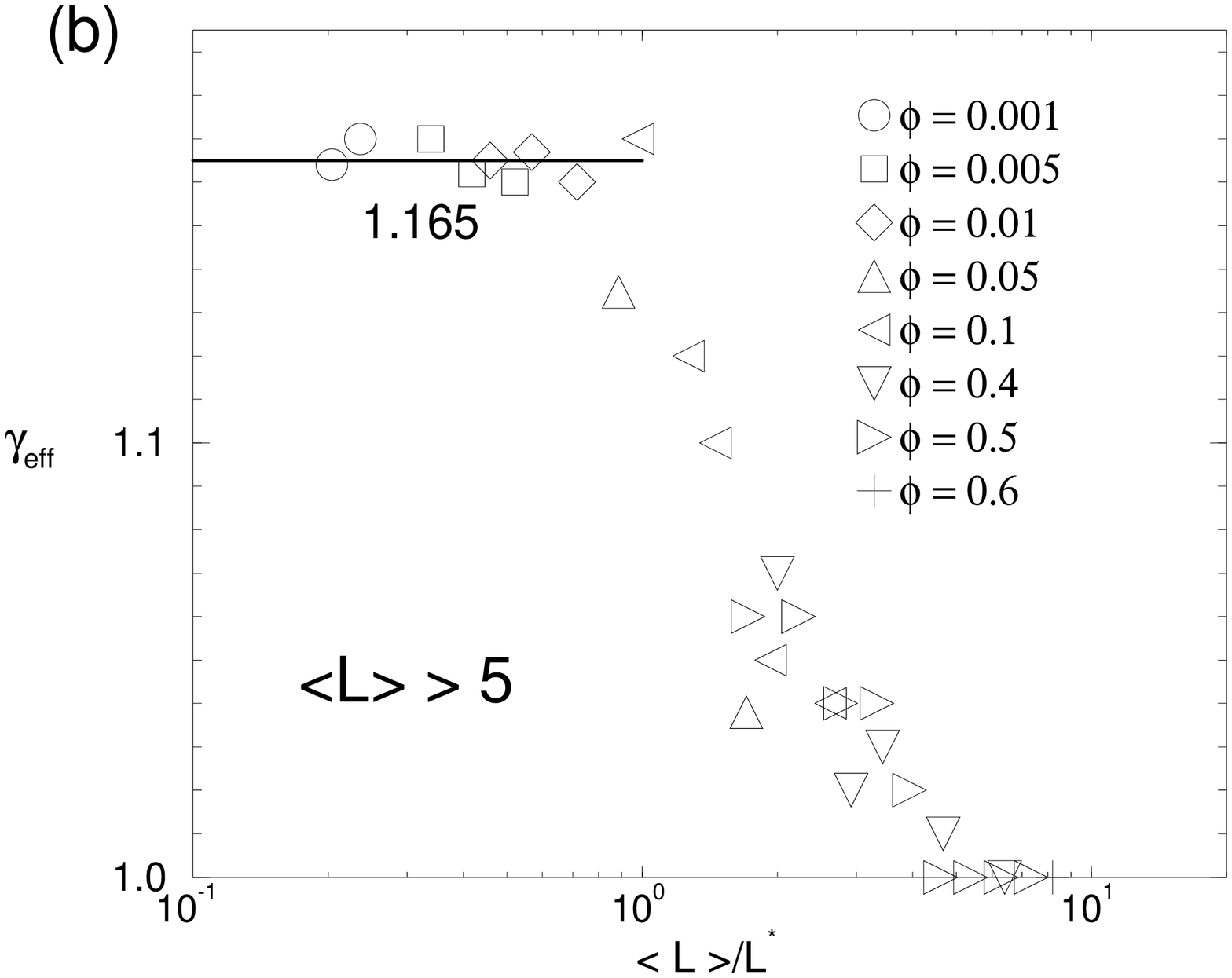,width=70mm,height=80mm,angle=-90}
}
\caption{Molecular Weight Distribution.
(a) Dilute limit confirming
$p(x) \propto \exp(-\gamma x)$ with $\gamma \approx 1.165$.
Insert: MWD for small $x$ for two configurations with $\phi=0.005$ (squares)
and $\phi=0.01$ (diamonds) both at $E=10$ in the dilute regime.
The prediction of Eq.~(\protect\ref{eq:p}a) is also given for
comparison (bold line).
(b) Effective exponent $\gammaeff$ versus the number of blobs per chain
$\Lav/\Lstar$.
\label{fig:MWD}}
\end{figure}

The normalized MWD curves, $p(x)$ plotted versus the reduced chain length
$x=L/\Lav$, likewise collapse perfectly on single `master' curve as shown in
Fig.~\ref{fig:MWD} for systems in the dilute limit.
With larger values of the reduced chain length $x$ one observes an
increasing statistical
noise since very long chains are more rarely encountered in the simulation.
Our results are qualitatively
consistent with the Schultz distribution Eq.~(\ref{eq:p}b) (bold line) and
agree with earlier studies of the MWD of EP in two dimensions\cite{YA2}
where the much larger value of $\gamma_{2d}=43/32$ strongly enhances
the effect. In three dimensions the difference is smaller yet detectable.
We have also
included the high-density prediction so as to emphasize the observed deviation.
As shown in the insert of Fig.~\ref{fig:MWD}a, our MWD in dilute
solutions is qualitatively consistent with the additional power-law
dependence $p \propto x^{\gamma-1}$ in the limit of small $x$.
(Note that maximum of the distribution is at
$x_{max} = (\gamma-1)/\gamma \approx 0.1$
where we have chain lengths of only  $L \approx 4$.)

In the semi-dilute regime (not shown) the normalized MWD shows an equally 
good data collapse as a function of $x$, 
confirming the prediction Eq.~(\ref{eq:p}a)~\cite{WMC}.
At intermediate densities, slightly above the dilute-semi-dilute crossover line,
a non-negligible fraction of chains are smaller than the blob size, and
being thus fully swollen, may exist within the network made up from the
chains of average size $\Lav$ without being seriously perturbed by the
interchain interaction. The molecular weight distribution crosses therefore
smoothly over from the dilute limit to the semi-dilute limit; detailed
data will be presented elsewhere~\cite{WMC}.
Here we elucidate this crossover by analyzing the effective
exponent \gammaeff\ found from fitting to the expected $\exp(-\gammaeff x)$
tail in the MWD at large $x$. 
The obtained values are plotted in Fig.~\ref{fig:MWD}b versus the 
number of blobs $\Lav/\Lstar$.
It is clearly seen that $\gammaeff \rightarrow \gamma=1.165$ and
$\gammaeff \rightarrow \gamma=1$ in the dilute and semi-dilute limits
respectively. 
In between we observe a smooth crossover for the effective exponent.
The error bars are around $\pm 0.01$, however, a much
higher accuracy is in principle feasible with our method~\cite{WMC}.
Note also that the heterogeneity index $I=\left<L^2\right>/\Lav^2=1+1/\gamma$, 
(the ratio of weight average and number average molecular weights) allows for 
an alternative determination of $\gamma$.
However, due to the strong contribution of the distribution at small $x$,
which is -- as mentioned above -- influenced by finite chain-length effects,
this method systematically overestimates the exponent~\cite{WMC}.

In summary, we have presented new simulation data on the statistics of
Equilibrium
Polymers. By use of an efficient Dynamic Monte Carlo algorithm, we are able
to study much larger mean chain lengths (combined with high densities) than any
previously examined. The present work unambiguously confirms the scaling results
for EP without rings, based on the classical behaviour of conventional quenched
polymers \cite{Cates88,Schaefer,Schoot}. In particular our results do not
support a value
of the semidilute growth exponent $\alpha$ larger than the classical one
($\alpha
\simeq 0.6$), as suggested previously~\cite{Schurtenberger} in explanation
of the anomalous
viscosity of giant micelles in nonpolar solvents.  We also find no evidence
for a possible {\em negative} exponent in the power law in Eq.(\ref{eq:p}),
as postulated in some treatments of the unusual diffusive behaviour in
giant micelles \cite{Bouchaud}.  The extremely large size of our chains,
allied to a careful analysis of the size distribution in the dilute limit, 
allows us to extract an accurate estimate of the self-avoiding
walk susceptiblity exponent $\gamma = 1.165\pm 0.01$. This compares with
the best renormalization group estimate 
$\gamma = 1.1615\pm 0.0011$~\cite{desCl}.

\vspace{0.5cm}

The authors are indebted to J.P.~Desplat and Y.~Rouault for valuable 
discussions and assistance during the present investigation.
This research has been sponsored by National Foundation Grant No. Int. 9304562
and by the Bulgarian National Foundation for Scientific Research under Grant
No. X-644. 
AM acknowledges the hospitality of the EPCC in Edinburgh (TRACS program).
JW is indebted to D.P.~Landau for hospitality
at the University of Georgia, Athens, USA. 


\end{document}